\documentclass[onecolumn,preprintnumbers,amsmath,amssymbn,reprint,nofootinbib,superscriptaddress]{revtex4}    

\usepackage[english]{babel}
\usepackage[utf8x]{inputenc}
\usepackage[T1]{fontenc}
\usepackage[normalem]{ulem}
\usepackage{color}

\begin{document}

\title{Obliterating Thingness: An Introduction to the ``What'' and the ``So What'' of Quantum Physics}

\author{Kathryn Schaffer}
\email{kschaf2@artic.edu}
\affiliation{ School of the Art Institute of Chicago,112 S. Michigan Avenue, Chicago, IL 60603, USA}

\author{Gabriela Barreto Lemos}
\email{G.BarretoLemos@emb.edu}
\affiliation{International Institute of Physics, Federal University of Rio Grande do Norte, Campus Universit\'ario - Lagoa Nova - 59078-970 - CP.: 1613, Natal/RN, Brazil}
\affiliation{University of Massachusetts Boston, 100 William T, Morrissey Blvd, Boston, MA 02125,USA.}

\begin{abstract}
This essay provides a short introduction to the ideas and potential implications of quantum physics for scholars in the arts, humanities, and social sciences.  Quantum-inspired ideas pepper current discourse in all of these fields, in ways that range from playful metaphors to sweeping ontological claims.  We explain several of the most important concepts at the core of quantum theory, carefully delineating the scope and bounds of currently established science, in order to aid the evaluation of such claims.  In particular, we emphasize that the smallest units of matter and light, as described in quantum physics, are not {\em things}, meaning that they do not obey the logic we take for granted when discussing the behavior of macroscopic objects.  We also highlight the substantial debate that exists within physics about the interpretation of the equations and empirical results at the core of quantum physics, noting that implicit (and contested) philosophical commitments necessarily accompany any discussion of quantum ideas that takes place in non-technical language.
\end{abstract}
\maketitle
\section{Introduction}
\label{intro}
This essay is a short introduction to some core concepts and philosophical problems associated with quantum physics. We are writing it to respond to, and to enhance, conversations about the meaning of quantum physics that are currently underway in contexts beyond the physics laboratory.  

{\em Far} beyond the physics laboratory.  We are two physicists who regularly work with artists and designers.  We increasingly hear from colleagues and students in these creative fields that quantum physics is an important source of ideas for their work, even though they may have never taken formal physics courses.  In our personal experience, the ideas of quantum physics seem to be undergoing vigorous ``cultural processing'' in this historical moment, largely beyond the gaze of professional physicists.  

``Cultural processing'' is our own term.  It is meant to loosely encompass anything that people do with the theories, empirical results, narratives, or methodologies of a scientific field that takes place outside the central institutions and practices of scientific research.  It is an umbrella term for the many heterogeneous ways that some ideas from the world of science end up having meaning outside their original contexts.

Many scientists simply object to the idea that scientific ideas could have meaning outside their original contexts.  We do not.  For one thing, any time that scientists themselves attempt to translate scientific ideas for a public audience, they are engaging in a form of cultural processing.  The drive to share scientific ideas, and to re-express them in non-technical language, speaks to the potential for science to have meaning beyond a research paper or a technological application.  It would be ridiculous to think that big ideas like relativity or quantum entanglement would have no importance beyond the lab; their non-technical relevance (including the potential for shaping worldview) is a part of why many scientists pursue, and share, their research in the first place.  

We also believe that questions of the meaning of scientific claims already are (and should be) open to the input of non-scientists.  Enabling interdisciplinary exploration into the philosophical, metaphorical, and generative potential of quantum physics concepts is one of our aims with this essay. 

In the art and design world, we see a particular demand for greater understanding of quantum physics spurred by the influence of a single interdisciplinary feminist scholar, Karen Barad\cite{Barad}\footnote{Barad is faculty at the University of California, Santa Cruz.  Her book \cite{Barad} is widely read in the humanities, in science studies, and in the arts, although it is interesting to note that she is almost entirely unknown among physicists.}.  Barad (who herself was trained as a physicist) has opened up entirely new communities of interest in quantum phenomena.  You do not need to be familiar with Barad's work to understand the rest of this essay, but we will refer to her as a central example of someone who takes the ideas of quantum physics to be deeply meaningful outside of the laboratory setting.  Barad believes that the facts of quantum physics are so philosophically important that they wholly change how we should think about people, relationships, subjectivity, objectivity, nature, culture, and scholarship itself. Citing Barad, scholars in the arts, humanities, and many interdisciplinary fields now write about the ``observer effect'' and ``entanglement'' -- technical physics concepts -- in work that has a distinctly social or political (that is, not primarily physics-based) emphasis.  

In the social sciences, quantum concepts have also gotten a recent boost due to the work of International Relations scholar Alexander Wendt\cite{Wendt}. Wendt thinks the social sciences have been led astray by implicit assumptions of a mechanistic and deterministic (``classical'') universe.  In contrast to Barad, however, he also engages in some provocative speculation beyond currently established science. To Wendt, the unresolved problem of the nature and origin of consciousness is a critical barrier towards progress in any field attempting to study humans (thinking beings with subjective, conscious experiences) scientifically.  Wendt believes quantum physics will ultimately be important for explaining consciousness.  Thus, Wendt seeks more from quantum physics than the inspiration for new ideas.  He believes it may be critical to the social sciences because it may be the underlying science of how humans think and behave.  

That could be true.  Or, it could be false.  Or it could be a piece of a more complex truth.  At the moment, we simply do not know, since the scientific understanding of the mind, brain, and conscious experience is still far from complete.  One reason for noting Wendt's work here is to emphasize the importance of empirical tests to evaluate any claims that take steps beyond what we have already established scientifically, including claims that quantum physics has something to do with consciousness. Wendt acknowledges that his argument includes a gamble, and that future science may falsify some of his claims.  

Unfortunately, the cultural processing of quantum physics more broadly also involves quantum pseudoscience proponents, who promote (and profit from) unsubstantiated and unscientific claims that quantum physics has something to do with human thought, spirituality, or health\cite{Pessoa}.  These forms of pseudoscience, which we encounter surprisingly often in conversations with students and colleagues, are problematic because they are at odds with the established knowledge and methods of science. 

To be clear, speculating beyond the limits of current science is perfectly fair game for anyone, and such speculations need not be responsible to the knowledge and methods of science in every context (for example, in science fiction).  Yet we think it is important to be able to tell the difference, and to respect the things that set scientific knowledge apart from other forms of thought or belief. Any claims of the sort ``A physically causes B'' are explicitly the kind of claims that should be substantiated by rigorous scientific research, which includes experimental verification.  
Such substantiation, generally speaking, is absent for pseudoscientific claims.  

The specific quantum healing claim that the act of conscious thought can directly cause changes in the external world is problematic at the outset, because science has no complete model to describe what conscious though {\em is}, much less to model ways it might exert a causal influence. If the proposed mechanism for such influence is claimed to be "direct," on the basis of some quantum physics effect (not through complex and multifaceted psychological and social mechanisms), that claim is not based on any empirically substantiated science; no such mechanism is proposed or described within the field of quantum physics as it currently stands.  Even if we take the claim as speculative, any possible mechanisms for physical causation need to be evaluated for consistency with everything else we know about how the world works. Precisely because we have such a sophisticated understanding of the forces involved in physical interactions (enabling a host of technologies from brain scanning to remote sensing), this is a high bar to clear. Any causal mechanism that is supposedly based on physics needs to be explainable {\em along with} and {\em in relation to} all the other physical causal mechanisms we already understand.  In other words, any proposed physical mechanism for quantum healing needs to be explained in the same framework that explains MRI scans, thermal imaging, and all the other mechanisms we already know for physical communication with and about the body.   

This is all a long way of saying:  if it sounds too good to be true, it probably is. Quantum physics is not a trick or a way of evading the ordinary rules of nature.  Quantum physics has some amazing implications, but it is very much grounded in the physical and the possible, describing processes that are going on in atoms, in computer chips, in lasers, and in nuclear bombs.  Powerful stuff, but all well-understood, empirically founded, technologically activated physics. 

In the next several sections, we will say more to define the scope of quantum physics, discuss some core representational and philosophical issues, and describe some of its key empirically-founded insights.  Throughout, we will directly address possible philosophical conclusions one could draw from quantum physics, as well as try to clearly draw the line that divides science from pseudoscience.

\section{The Scope and Form of Quantum Physics}

To a physicist, {\em quantum physics, quantum mechanics}, and {\em quantum theory} all refer to the same thing:  our physical theory for phenomena on very small size scales (comparable to, or smaller than, the individual molecules and atoms that make up materials around us)\footnote{In some contexts, scholars may differentiate these terms.  In the physics context, however, the terms are interchangable: any could be used as the title for an introductory course or textbook covering the same material.}.  A {\em theory} in physics is much more than a set of ideas.  It expresses quantitative relationships between things we can measure, which means it involves equations. There are some comparatively clear criteria for determining whether a theory in physics is a successful one or not.  To be successful, it must be able to account for experimental results that have already been obtained, and, more importantly, be able to predict the outcomes of future experiments. Successful theories become integrated into the working practice of physics. A central part of that research is to explore the consequences and applications of the theory, as well as to continually develop more rigorous tests that probe its scope and limitations.  Any robust and repeatable observation that is in conflict with the theory may require a revision to or replacement of the theory. If this does not occur, and evidence in support of the theory grows, it becomes even more deeply ingrained in how we do and think about physics.

Quantum theory is one of the most successful physical theories ever developed.  Since its foundation in the early twentieth century, this theory has been tested through decades of rigorous experimentation. Its track record of accurate prediction is astonishing, allowing countless technologies to be designed on the basis of those predictions.  Every digital camera and smartphone in the world is a testament to the success of quantum theory, as is every nuclear power plant.  Quantum theory is so fundamental to our understanding of nature that it underlies entire fields of scientific research (e.g. chemistry).  Quantum physics is thus not at all speculative.  It is not a form of philosophy, and it is not something that is principally expressed, or employed, through verbal language.  At its center is a practical toolkit of equations that technical experts use to model, understand, and design a wide range of structures and phenomena that involve light, electrons, and atoms.  

One of the most remarkable lessons from 20th century experimental physics is that light and electrons and atoms require a very different kind of description than macroscopic objects and events.  There are {\em two} important points here. First, nature is not the same on every scale. On small scales, bits of matter move and interact in completely different ways than bodies in a room or planets around a star.  Second, we discover that the equations we need to describe and predict those motions and interactions are of a completely different character, too.  

If you are not used to thinking about how equations relate to things in the world, it might be hard to imagine what we mean by equations having a ``different character.''  It might also be hard to believe that talking about equations is important in a supposedly non-technical introduction. However the core content of quantum physics is expressed through equations, and one of the key points we want to make is that there is ambiguity any time you try to translate those equations into words.  This opens up philosophically interesting (and possibly problematic) territory, so it is worth highlighting.      

Whether we are talking about macroscopic or microscopic phenomena (large or small scales), physics deals with things like motion, interactions (like collisions), forces, causality, and changes in physical systems over time. To take an especially simple example, imagine that some asteroids in deep space collide and a chunk of rock goes hurtling away into space.  The role of a physical theory is to do something like provide an equation to describe the motion of that rock.  The equation will contain symbols that stand for the measurable properties of the physical system.  The mathematical relationships between symbols in the equation serve as a model for the physical relationships between the properties themselves.  The model (equation) can be manipulated to learn things about the physical system, like where the rock will be at a future time. 

If we instead consider a single electron flying through empty space, we are in the realm of quantum physics and we need to use a different equation.  Let's compare the equations we would use in the macroscopic and microscopic cases, to talk about some of their differences (don't worry - you do not need to do any mathematics to follow the discussion).  

An equation we could use to describe the motion of a rock hurtling through empty space might look like this:
\begin{equation}
x(t)=x_0+vt.
\end{equation}
Each symbol in this equation has a meaning that we can express in plain language\footnote{Note that, in this equation as well as any other used in physics, it doesn't matter what the symbols actually are as long as we know what they represent, or how to use them.}. The symbol $x(t)$ represents the position of the rock, along its direction of motion, as a function of time.  The symbol $x_0$ represents the position of the initial asteroid collision that sent off the rock at some speed.  The symbol $v$ represents the speed or velocity of the rock as it hurtles through space.  And $t$ represents the time that has elapsed since the rock started moving.  

Now, in contrast, here is an equation we could use to describe an electron freely traveling through space:
\begin{equation}
\Psi(x,t)=\frac{1}{\sqrt{2\pi\hbar}}\int^\infty_{-\infty}\phi(p)e^{ipx/\hbar}e^{-iEt/\hbar}dp.
\end{equation}
You will immediately notice that it looks {\em a lot} more complicated.  While the first equation used only algebraic operations, this one depends on calculus (note the long curly integration symbol) and has imaginary numbers in it (the $i$), which is why quantum physics is not usually taught early in an individual's education.  The most important point we want to make about this equation is that, in contrast to what we did with the previous equation, we cannot simply tell you what each symbol in the equation means.  Arguably, nobody can.  Yes, all of the terms in the quantum equations are well defined in the sense that physicists know how to use them operationally, and how to relate them to experimental results.  Yet we have problems when we try to say anything in words about why the equations work the way they do, or what underlying structure of reality yields the behavior they predict.

In classical equations governing macroscopic physics, we treat measurable physical properties as if they have definite physical reality and definite relationships to one another.  Those definite properties have a direct one-to-one correspondence with the terms in the equation.  The term $x(t)$ means the position of the rock at a time $t$, which is understood as a well-defined property of a thing that we could measure in any way we pleased.  The equation then expresses a relationship between that property and other properties (the velocity and starting location) that are similarly understood to be well defined and independent of the method and sequential order of a set of measurements.   All the ways we ordinarily talk about properties and relationships in everyday language apply.  

In the quantum case, nothing in the equation stands for anything analogous to $x(t)$.  Specifically, the term $\Psi(x,t)$  -- the wavefunction for the electron, which is what the equation describes -- is not necessarily a property of anything in the physical world.  It functions as a tool.  $\Psi(x,t)$ encapsulates statistical predictions for the probability of finding the electron in a certain place at a certain time.  In quantum equations, we explicitly lose any references to properties such as position as if they are clear, definite, facts about the world. Moreover, nothing in the mathematics tells us what underlying physical mechanism leads to the need for a probabilistic description in the first place.  That is, it does not tell us what the electron itself actually is, what its properties are, nor how it ends up behaving in this odd way.

Measurement also matters in the quantum physics mathematics in a way that it does not matter for classical physics computations.  When we manipulate the equation for the rock, we do not need to take into account whether we are thinking about measuring the velocity first or the position first.  When we manipulate the quantum physics equation, we have to explicitly account for the types of measurement we might make, and their order. The order of measuring the properties of the electron matters.  In our mathematical model, each type of measurement actually changes the wave function itself.  

Thus the way quantum equations function is simply different.  They do not have terms in it that relate to simple, nameable {\em things} nor to permanent and independently knowable {\em properties} of those things.  Quantum theory gives us a probabilistic description of many possibilities.  Nobody knows for sure why these recipes work, nor how to talk about the relationship between the mathematical operations and the underlying physical nature of the electron itself.  

The point is that it does work.  Both equations given in this section work, just in different contexts (macro versus micro).  Both equations are idealizations, and real scenarios often require more complex versions (to take into account forces due to gravity or due to other particles, for example).  But, when the conditions are close to the ideal, the equations function predictively and descriptively, in their different ways.  Macroscopic physical reality can be described with equations that have nameable things and well defined properties.  Microscopic reality requires probabilistic equations with a less direct relationship between the symbols and anything we can simply name or define.  

\section{The Boundary Between Theory and Interpretation} 

We seem to be stuck accepting that quantum mechanics equations are just different.  Most importantly, they make no unambiguous references to the structure or form of physical reality prior to specific measurements. Quantum physics does not tell us what the electron {\em is}, or what the wave function {\em means}.  To go the extra step of assigning words to the things that are represented in this scheme, we have to pick a particular {\em interpretation}. 

The interpretation is a set of philosophical commitments associating the terms in quantum mechanics equations, and the phenomena observed in laboratories, with specific meanings.  This step is necessary if we want to say things like ``in quantum physics an electron is...'' or ``when the electron went through the apparatus, what happened was....''  The point is that such sentences will come out differently if we make different interpretational choices.  

Thus, and this is the punchline of this section (and a key punchline of the whole essay) {\em there is no single quantum ontology}\footnote{``Ontology'' is a word from the field of philosophy that refers to a theory of ``what is'' in the universe.  It is often contrasted with ``epistemology,'' describing a theory of how we know about things in the universe.}. A quantum ontology would be a scientifically supported way of answering questions like these:  What is actually going on with the electron flying through space?  Is the electron itself actually ``spread out,'' physically embodying many possibilities at the same time? Does $\Psi$ correspond to a real physical thing, or does it capture something only about what is knowable about a situation?  Could there be multiple versions of reality -- multiple universes, even -- in which the electron is in all of the different places that are expressed as possibilities in $\Psi$?  Is the randomness we observe -- the need for a probabilistic description -- something fundamental to the universe, or an expression of limited knowledge?  

All of these are questions about the structure of physical reality on the quantum scale, and none of these questions can be answered unambiguously by the physical theory itself.  While quantum theory is wildly successful and well proven as a tool, it leaves open major questions about how the universe actually works. Although we will allude to a few specific interpretations in this essay, there are nearly as many interpretations of quantum mechanics as physicists working in the field.  
A group of quantum physicists might include individuals who subscribe to a variety of different interpretations \cite{interpretations} such as, for example, the Copenhagen interpretation \cite{copenhagen}, QBism \cite{qbism}, the Information-Theoretical interpretation \cite{information,Bub}, the Relational interpretation\cite{Bitbol, Rovelli}, the Many Worlds and Everettian interpretations\cite{manyworlds}, Bohmian Mechanics (also called de Broglie-Bohm or Pilot Wave interpretation)\cite{bohm}, Consistent Histories\cite{histories}, the Time Symmetric interpretations \cite{retrocausal, aharonov,Wharton07}, or the Objective Collapse Models\cite{collapse}.  Or, none of the above.

Most non-technical writing about quantum physics does not emphasize this point.  Authors typically pick an interpretation and explain quantum physics from within that framework. It is hard enough explaining the weirdness of quantum physics within a single interpretation, much less trying to explain that everything could be completely different if we picked another.  But if you are interested in asking about the meaning of quantum physics ``outside the lab,'' we think that it is important to acknowledge that there is no consensus on the meaning of quantum physics ``inside the lab.''  

This point is relevant, for example, if you are reading Karen Barad. She takes quantum ontology as the starting point for rethinking all ontology (as well as epistemology and ethics, in fact). To do so, she must commit to an interpretation in order to have a quantum ontology to start with.  In her case, she picks an interpretation that is widely favored among physicists (for historical and cultural reasons, not because there is any evidence supporting it).  In this interpretation\footnote{Which is known as the ``Copenhagen interpretation,'' referencing the place where it was mostly developed by Niels Bohr.}, something like an electron is treated as a fundamentally indeterminate entity prior to measurement.  This means that it does not have well-defined properties (like location, or velocity) until it is measured.  Measurement (which need not be measurement by conscious humans, but could be some form of interaction with an environment as well) creates definiteness.  Barad takes this as a fundamental fact of how the universe operates:  definite properties, and definite things, only emerge through interactions.  The notion that anything, on any scale, has a persistent and well-defined identity, is thus called into question.  

Our point is that you might end up with quite different philosophical conclusions if you started with a different interpretation of quantum physics.  In some, entities like electrons (and everything else) have perfectly well defined properties.  To account for quantum phenomena, we know that it is impossible to have complete access to information about those properties (otherwise, different forms of equations, more like the classical case, would work).  But the existence of definite things and definite properties is not ruled out by any established quantum physics.

A final note for this section on interpretation is that the lack of a single clear interpretation does not mean that the nature and structure of the universe is a philosophical free-for-all.  There are many speculative or imagined ideas about quantum physics that are simply inconsistent with empirical facts or the scientific method (like many quantum healing claims, as we mentioned in the introduction).  What an interpretation of quantum physics deals with is the meaning we assign to (a) terms that show up in equations or (b) phenomena that are observed in well-controlled, repeatable physics experiments, like the kind that are described in peer-reviewed research publications. If an author or speaker claims to discuss the physical, causal implications of quantum physics and there are no equations or rigorously-performed quantum physics experiments involved (at least in the background), it is not actually about quantum physics, in any interpretation.  Period.

\section{Core Ideas of Quantum Physics}

We have argued that there is no single framework for discussing quantum phenomena through language.  Short of listing empirical results and providing equations and recipes to predict them, there can be no ``interpretation-free'' description of the microscopic world\footnote{Arguably, even the choice of the way that the equations are constructed is linked to interpretation, although ultimately all successful formulations of quantum physics need to be mathematically equivalent wherever they link to descriptions or predictions of well-established quantum phenomena.}.

Nevertheless, an equally important point is that no matter what physical reality corresponds to the equations of quantum physics, it is a weird one.  Weird, meaning inconsistent with what you would expect based on macroscopic experience\footnote{This point is argued nicely in the book \cite{ontology}, where the author argues that, while there is no single quantum ontology, all possible interpretations of quantum physics are philosophically significant. See also \cite{Pessoa} for a discussion on the philosophical implications of the different interpretations.}.

What we want to do in this section is call out some of the core facts of quantum physics that, regardless of how we interpret them, are in conflict with the intuition and experience we have on the macroscopic scale.  We will use a mix of analogies, hypothetical examples, and a little bit of our own invented language, in a deliberate attempt to avoid some of the most common (and tightly interpretation-bound) pedagogical constructs, letting us hopefully emphasize the interpretation issues more clearly.

\subsection{Quantization}
\label{quantization}

Quantum physics gets its name from one core fact:  at the smallest scale, nature is ``digital'' not ``analog.'' Think of the difference between a digital and an analog clock. In the digital clock the smallest ``chunk'' is a second, whereas an analog clock runs continuously.  The second-hand can be in between two one-second tick-marks on the clock dial. Physical entities like matter and light come in smallest chunks, like the bits and bytes of digital information.  So do properties of those physical things, like their motion energy, or their electric charge.  {\em Quantum} (plural {\em quanta}) is the name given to an individual chunk.

This fact, by itself, means that the rules of the game are different on a microscopic scale. Basic physical quantities like energy and momentum can only be exchanged in certain specific quantum units.  This imposes constraints on the interactions that are possible among quanta of matter or light.  In physical interactions, quanta can only exchange energy or momentum in whole quantum ``chunks,'' and never in smaller amounts.  In the macroscopic world, it is as though a tea kettle heats up by gradually warming up from zero to the boiling temperature, spending at least a tiny moment at every temperature in between.  Microscopically, there are specific steps to any such process, and there simply is no ``in between.'' 
\subsection{Non-thingness}
\label{nonthingess}

Electrons and other particles that make up matter are themselves quanta.  Photons (individual ``chunks'' of light) are quanta.  So what are quanta?  Well, that's where we hit the interpretation problems described in the previous section.  There is no single way to talk about what they are, so let's focus instead on what they are not.  The single most important idea to grasp about quanta is that they are {\em not things}.  This is at the heart of the radical weirdness of quantum physics.

Let's define what we mean by {\em things} so that this statement gains some weight.  By {\em things} we mean {\em objects} or materials that operate by the familiar rules and logic of the macroscopic world.  Examples include coffee, cats, cars, carpets.  {\em Things} take up space.  If they move, they do so in a continuous way along a single trajectory in space.  While a thing may have a large physical extent, we never talk about it being in two completely different places at the same time.  Things also cannot jump instantly from one place to another.  They have physical properties, like their size, location, or speed of motion.  Those properties may change over time, but at any single moment, the properties have definite values that can be used to describe the thing in question.  Things continue to exist when we're not looking at them.  If they are created or destroyed, compounded or broken apart, there is a single narrative we can use to describe what happened.  

None of the statements we made about things, above, can be applied in a straightforward way to describe how quanta work.  To get a taste for this, there is an experiment you can try at home. Take boxes and lay them out making a pentagon shape. Now take some red balls and some blue balls in your hand and place exactly one ball (of either color) in each box. Call a friend into the room and ask them to open any two adjacent boxes. Is it possible to arrange the balls (one in each box) such that your friend will always find one red ball and one blue ball, for any two adjacent boxes they decide to open?  If the balls were not \textit{things} but \textit{quanta}, then it would be possible to arrange such a scenario with different sequences of measurements.\cite{Liang}

Of course, it isn't simple to say ``instead of following macroscopic behavior, quanta work like this...''  because while the empirical facts are well established, the words we would use to describe them are tied to specific interpretations. Given a single scenario involving an electron in a laboratory, one physicist might be comfortable describing its behavior by saying ``the position of an electron is intrinsically undefined, all we know is that it behaves as if it were in many places simultaneously.''  Describing the same physical scenario, another might say ``an electron is a spread-out entity that does not have a single location.''  Or, ``the electron always has a single, definite location, but knowledge of that location is deeply impossible.''   Or even, ``many parallel universes exist, and in each of those, a copy of the same electron exists at a different place.'' These are a just a few of the many radically different (and quite radical) statements linked to different interpretations of the electron's non-thingness.  The equations describing all of these statements are the same in each case.  The observed behavior of the electron is the same in each case.  The electron's departure from everyday physics is the same in each case.  But the words, and the worldviews that accompany them, may be quite different. 

Since we view non-thingness as a central feature of quantum physics, we would like to help you to build some intuition for it through analogy.  Humans work all the time with abstract concepts that have some non-thing-like behaviors. For example, money\footnote{We will use the money analogy casually here, but it is potentially quite nuanced. As we were writing this essay, we had some productive exchanges with David Orrell, author of \cite{orrell}, who argues that money shares many properties with the quantum systems studied by physicists, and perhaps should be modeled with a similar type of mathematics.}.

To explore money as an analogy for conceptualizing quanta, imagine that you have some dollar bills in cash and you deposit them into your bank account using an ATM machine in Chicago.  You put real physical money into the machine at a specific location.  But you know that as soon as the machine counts the bills and credits your bank account, any meaningful relationship to tangible dollar bills is lost.  When you held the dollar bills, the money had a well-defined place:  it was in your hand.  Once you deposit in the bank, where exactly is it?

Sure, the ATM creates a computer record, and that computer record is located somewhere (probably duplicated in many places).  Yet it doesn't seem right to say ``the money becomes bits stored in a computer.''  If the whole transaction were recorded on paper instead of bits on a computer, it would still be the same money.  Money-in-the-bank is an abstract concept that does not necessarily depend on the form of any particular record we use to keep track of it.  

This abstract concept of {\em money in the bank} or {\em a dollar you own} is a concept that behaves much like quanta do.  Dollars that you own do not always have a well-defined trajectory in space, and we cannot always sensibly ask where they are at any given moment.  Suppose you deposit some dollar bills into the ATM in Chicago, and later fly to L.A..  You can withdraw your money from an ATM there.  Would you say the money was somehow in L.A. before you went there?  How did it ``know'' you were going to L.A. and not New York?  If you had chosen to go to New York, you would have been able to withdraw it there.  In a sense, then, your money is equally present anywhere that is connected to the same bank network, and where you find it at any given moment depends on where you initiate a bank transaction.  This is a lot like the way that a quantum lacks a well-defined location in an apparatus, until it is measured.  

Along these same lines, the money does not need to pass through points in between two locations where you enact transactions with it.  We would not say that between your transaction in Chicago and your transaction in L.A. that the money must have been in a city like Denver, somewhere in the middle.  Of course, if you go to Denver, you can make your money be there by initiating a bank transaction there instead.  But would it have been there without you?  Would it have been in any of the cities along any route from Chicago to L.A.?

The lack of definite trajectory in this example is similar to the behavior of electrons and photons and other quanta.  It is a weird comparison, because money is an invented abstraction, and electrons and photons and other quanta are constituents of the touchable, viewable physical world.  Yet, the intuition you have for the way money works is a useful start for grasping the non-thingness of quanta.

One useful feature of the analogy is the way that your transactions play an active role in determining where your money is.  In the physical world, if someone or something interacts with a quantum, it changes the quantum's behavior. This is known as the ``observer effect,'' although it does not necessarily require a conscious observer.  Consider a quantum like an electron that is sent through an apparatus in which it can travel multiple paths.  We discover that it does different things depending on where (on which path) we place our detectors.  That is, the act of detecting the quantum actively changes what it does.

One idea that the money analogy does not quite capture is that the mere existence of multiple possible physical trajectories can affect the outcome for individual electrons passing through an apparatus.  This is a bizarre thing, that is not at all true for money in the bank.  The mere presence of a path through Denver as one route from Chicago to L.A. does not change the behavior of your money.  In a physics experiment, different outcomes will happen if more paths are present, even if every measurement only ever shows it on one single path.  This is exactly the kind of experimental result that leads to the interpretative disagreements we described before:  is there a guiding force that makes the electron act as if it were in many places at once?  Is the concept of {\em location} just something we can't use with electrons, when we are not actively observing them?  Are there many copies of the same electron simultaneously taking all possible paths in many universes?

In the end, even if some analogies can help provide some intuition about what we mean by non-thingness, we are likely to hit dead ends with every analogy that uses words or familiar everyday concepts.  The familiar, and the everyday, is rooted in the macroscopic world, and the microscopic world simply plays by different rules.

\par 
\subsection{Randomness}
All quantum phenomena display randomness.  

We encounter randomness everyday on a macroscopic scale, but the randomness in quantum physics is of a different character.  For example, consider flipping a coin and obtaining a random result, heads or tails.  The way this differs from quantum randomness is that in macroscopic random events there are knowable (at least in principle) reasons why a particular outcome occurs.  You could make a movie of the coin flip, analyze the air currents, and reconstruct how the exact finger motion and trajectory of the coin through the air resulted in it landing heads-up.  In other words, we can construct a single coherent narrative of the coin, from the moment it was thrown to the outcome of the experiment.  It may be challenging in practice to predict or fully analyze the outcome of a coin flip, but it isn't impossible in principle. 

On the quantum scale, predicting or fully analyzing the outcomes of random events is impossible, even in principle.  (Well, to be fair, there are some disagreements about how far to go with the ``even in principle'' statement, which we'll explain in a moment).  One issue is that there is no way to continuously measure  (``take a movie of'') a random quantum process without physically interacting with it and affecting the outcome of the process.  There also is no single coherent narrative describing a quantum process leading to the prediction with certainty of an experimental outcome.  The logic that quantum random events follow in physics experiments is inconsistent with the idea that the entities have well-defined and knowable reasons for any given outcome.  

A quantum system analogous to a coin flip might be an experiment in which we send quanta through an apparatus and then measure a certain property that has a 50\% probability of having one value (call it ``A'') and a 50\% probability of some other value (call it ``B''). For example, it could have 50\% probability of facing up or facing down.  Between the beginning and the end of the experiment, was it facing up or down?  Was there some cause or reason responsible for an individual quantum ending up facing up or facing down?  Is that reason something that we could know?  We cannot answer these questions in the same way that we can for a coin flip, but to say more in words about what {\em is} going on, we have to take on a particular interpretation\footnote{The interpretation options described in this section are loosely related to the interpretation options presented previously, but different interpretations can take a mix of stances on different aspects of quantum behavior, like randomness.  There is such a large set of possible interpretations of quantum physics that we have decided not to attempt to enumerate or name any particular subset, but just to attempt to illustrate some of the differences as they apply to an individual concept like randomness.}.

For example, one interpretation of quantum randomness says that the quantum does not have actual properties until measured.  This is taking non-thingness to the extreme, to say that definite properties only exist in certain moments, like measurements, and not in the moments in between.  In this interpretation, randomness is truly fundamental, and no story or set of reasons can explain why quanta manifest as they do in any individual case.

Another interpretation is to say that the quantum does have properties between measurements but to know them would require knowing everything about the entire universe.  In this interpretation, there is a story that explains why the quantum ends up manifesting in a particular way, but that story potentially involves what is happening billions of light-years away.  Does that make it unknowable in principle?  We could debate what we mean by ``in principle'' and land on different sides of the argument,  but it certainly involves a different scale of unknowability than the practical unknowability of the outcome of a coin flip.  

And yet a third direction of interpretation says that the quantum does have properties between measurements but to fully know and characterize them we would need to have access to many worlds in which every possible outcome is equally real.   (And, of course, there are yet other interpretations that say other things).  

Again, we all agree that quantum (microscopic) randomness isn't the same as macroscopic randomness, because that's what experiments show us.  But when we shift to trying to explain what that means, every rigorously supported option has dramatic consequences in terms of how we think about physical reality and knowability. 

With all that said, this is another place we need to caution against over-reading the implications.  The fact that randomness is a seemingly incontrovertible aspect of fundamental reality does not mean we live in an ``anything goes'' universe, or that highly precise predictions are impossible.  Quantum randomness is built into the equations of quantum theory.  While those equations can only make statistical predictions, the statistical predictions are of very high quality.  We have to know where electrons will go, to high accuracy, when we design technologies like computer memory.  We have a great deal of knowledge about what quanta will do in most situations that quantum physics addresses, it is just knowledge that pertains statistically to the behavior of many quanta as an ensemble rather than exactly predicting the behavior of each single quantum. 

\subsection{Entanglement}

The final quantum oddity that we want to highlight is {\em entanglement}.   Entanglement is a term for a way that quanta can have fixed and definite relationships to one another while still individually showcasing the same deep quantum randomness.  In a sense, certain relationships themselves become more definable than the things doing the relating.  

To set up an illustration of this concept, first imagine you have a pile of identical coins.  You take half and you give half to a friend who then leaves town.  You agree that you are both going to do a little coin-flipping experiment and record your results.  You toss each one of your coins one at a time and record whether you get heads or tails, writing down all the results in a sequence.  Your friend does the same.  

You later compare your results and find that even though both of you saw apparently random sequences of results, there was a perfect match:  every time you obtained heads, they did too, and every time you obtained tails, they did too.  Could this ever happen in the ordinary, macroscopic world?  Well, yes.  We could imagine that the coins internally contained some complex system of microchips, clocks, and weights, pre-programmed to execute identical sequences even at a distance.  No matter how odd the correlation, we can always concoct the perfect conspiracy theory to explain what we see, even though of course ordinary coins would not be expected to show this kind of random-but-connected behavior.

Empirically, we find that quanta are capable of acting like these fictional coins, showing long-distance correlations despite behavior that is random.  This is the phenomenon known as entanglement.  The difference between entanglement and the coin example, though, is that quanta are {\em not things}. With quanta, we lose the ability to invoke ``conspiracy theories,'' because we cannot construct any single clear story about what they were doing prior to our measurement.  Experiments called Bell tests confirm that quantum correlations are capable of persisting even when all possible conspiracy theories are ruled out by the absence of well-defined quantum properties prior to the specific measurements we choose to make. (For a non-technical explanation of the idea behind these experiments see \cite{cakes}). 

Entanglement does not happen for coins, but it does happen for quanta.  It is not magic, in the sense that it is a feature fully described in the mathematics used by physicists.  But it is certainly dramatically different from the way that macroscopic reality works. What we see is that relationships among quanta can be preserved by nature despite the individual quanta behaving randomly.  Moreover, these relationships are maintained even when the quanta are separated enough that no physical signal (that is, one traveling at the speed of light or slower) could possibly reach from one to another in time to explain how they ``know'' about each other. Importantly, entanglement \emph{cannot} be used to instantaneously communicate  information from one place to another, because that would require a {\em causal} connection between the two quanta.  This is an essential point overlooked by most people.  Entanglement is a {\em correlation} and {\em not} a causal link of the kind that is necessary if you want to send a signal from one place to another. 

Because of the physical impossibility of causal signaling between two entangled quanta, quantum entanglement is more than long-distance correlation. In the macroscopic world, there are plenty of examples of long-distance correlation between random events.  For example, there may be correlations between the random fluctuations in stock market prices in the U.S. and Japan.  The difference is that stock markets around the world are physically connected by causal links that communicate information back and forth, unlike entangled quantum systems.  Imagine severing all phone and internet connections (and any other physical connection -- including global weather patterns and actual exchanges of goods and services) between the U.S. and Japan, as if the two stock market systems were operating on completely different planets.  In that case, we would expect the random fluctuations in each market to be uncorrelated, because they would have nothing to do with one another.  This thought experiment demonstrates that stock market correlations are a purely ``classical'' form of correlation, because they depend on the possibility of causal communication between the two systems. If you removed all possible causal connections, the correlation would disappear and the events would be independently random.  

Quantum entanglement is also more than the ability to have instantaneous knowledge about something at a distance.  Suppose you have one silver coin and one gold coin.  Without checking which is which, you slip one into your pocket and one into your friend's pocket.  Later, you can look in your own pocket and instantly know which coin your friend will find in their pocket, no matter how far away they are.  There will always be a perfect correlation between what you find and what they find, and you will  know something about their experience despite having no communication with them when they pull the coin from their pocket.  While this thought experiment shares some features with quantum entanglement, it too is a form of classical correlation.  The distinction here is that there is an unambiguous {\em fact of the matter} of whether you put the gold coin in your pocket or the silver coin. A third person could come check your pocket and they would know definitively which coin you had.  Their observation also would not change anything about the situation.  Quanta, on the other hand, behave as if there is no fact of the matter prior to measurement.  Entangled systems may have certain well-defined overall properties for the system as a whole, but we cannot treat the properties of individual quanta as definite or well-defined.  Moreover, a third person checking a quantum system will break the entanglement and destroy the possibility of further correlation.    

Quantum entanglement is odd, and different from macroscopic, classical forms of correlation.  It is extensively studied in laboratory experiments, but one important point to make is that the laboratory experiments that showcase long-distance entanglement generally require highly controlled environments.  Entanglement does happen constantly in nature, but maintaining entangled relationships for long times or over large distances requires that the entangled quanta do not interact with anything else.  

For this reason, quantum entanglement does not offer a particularly plausible mechanism for long-distance interactions among complex real systems like, say, human brains (or anything within human brains).  Any electron or photon in your brain is constantly interacting with the rest of the matter in your brain, and thus cannot maintain an entangled state with other quanta in the outside world.  Any time we are discussing complex structures of quanta (like complex chemical structures, or biological structures), long-distance entanglement effects are suppressed to the point of being irrelevant, simply because of constant interactions between quanta and their neighbors.
\section{Quantum to Macroscopic}

The previous section established that the microscopic universe behaves in fundamentally different ways than the macroscopic universe.  The macroscopic, everyday physical realm is the realm of {\em things} with definite properties and definite trajectories through space. Randomness occurs, but outcomes are still linked to a sequence of specific causes.  Physical relationships between physical objects are deeply tied to the physical objects themselves (we can't talk about the force exerted by the lamp on the table without there being a lamp and a table).  In abstract thought, we have concepts like money or love that may violate these precepts, but physical things do not. 

On the microscopic scale, quanta play by different rules.  Even though they are physically real entities, they defy description as {\em things}.  They show a different logic underlying their random behavior, and they demonstrate entanglement.  

How does one set of rules and behaviors transition into the other?  That is, how do the behaviors of quantum non-things build up to create the behaviors of macroscopic things?

Nobody knows.  

One possibility that physicists have considered is that it is a matter of the sheer size of the system. This would mean that there is a physical mechanism that acts on systems above a certain mass, inhibiting quantum weirdness, making it act like a macroscopic thing. Quite a few experiments are being carried out today that look for a potential size scale beyond which quantum features are suppressed due to gravity. No clear macroscopic to quantum boundary yet has been found, and many physicists believe there is no such boundary.

We do know that the more a quantum system interacts with its environment, the more thing-like it tends to become, because little particles in the air, or electromagnetic fluctuations in the environment, for example, can cause quanta to lose their non-thinginess extremely quickly. That is why most quantum phenomena only manifest in very strict laboratory conditions. The quanta must be isolated from the air in vacuum chambers and all interactions with the quanta must be delicately controlled. 

In any case, we can never directly access the quantum world. We can only know its effects on macroscopic measurement apparatuses. In a sense, we only access a translation of the micro-scale into the framework of macro things familiar to human experience. The quantum-to-macroscopic boundary is therefore a kind of ``language boundary.'' Just as with translations from one human language to another, there will always be some information that is lost in the process.  Since we cannot shrink ourselves down to the quantum scale, we may face fundamental limitations in understanding what the universe is like on the other side of the micro/macro divide.


Not only do we not know exactly how the quantum world actually works, but we don't know how the multiplicity of random quantum possibilities ends up translating to a single measured outcome, which is what we actually see.   How exactly does the measurement process affect the behavior of a quantum?  Why and how do the interactions with the environment or with a measurement apparatus make a quantum go from a non-thing to a thing? Does this happen instantaneously? Is this ``transition" from non-thing to thing more like an illusion, or does it correspond to a distinct change in the rules of physics?\footnote{In the many worlds interpretation of quantum mechanics everything behaves with quantum weirdness, even us. When we observe a quantum (via some measurement apparatus) we actually become entangled with the measurement apparatus and with the quantum. In the different branches of the universe different versions of us will see different measurement results. This is an example of an interpretation in which our experience of a different set of rules for the macroscopic realm is more like an illusion. } 

As you can see, there are many open questions about how the rules change from micro to macro.  That {\em something } changes is, however, a simple empirical fact. And it is an important point to remember when talking about how quantum physics might relate to the human realm.  Even the tiniest dust grain you can imagine has enough quanta within it, and is in such a constant state of interaction with its environment, that it loses quantum behavior.  That tiniest bit of dust is a {\em thing} with a definite place and definite physical properties.  Even if atoms within the grain of dust may, at individual moments, experience entanglement phenomena with each other or with their environment, the dust grain as a whole is not meaningfully entangled with anything else. If it drifts randomly in the wind, that random behavior is of the macroscopic variety, amenable to a narrative description in terms of cause and effect.  The equations a physicist would use to describe the grain of dust are simple, classical equations.

This is an important point to make because most of the things we care about in our everyday lives as humans are much larger than a grain of dust, and thus are even farther away from the quantum scale.  People, notably.  As far as physics is concerned, people are distinctly macroscopic entities, displaying none of the behaviors that quanta display, even if the quanta within our own bodies are busily doing their own thing in their full strangeness.  The details end up being irrelevant on our scale:  even if there is technically some entanglement that occurs between the outermost electrons of atoms in the layer of dead skin when my hand touches yours, this has no measurable or perceivable consequences of any kind for either of us.  It is a curiosity of the natural world that it occurs, but likely that's all there is to say about it.   

\section{So what?}

Alright, quantum physics is strange (or at least seems strange, to organisms adapted to function in the macroscopic world).  But so what?  What, if anything, does this have to do with everyday life, scholarship in other fields, or the problems and questions that we face as humans? There are a few ways you might answer this. 

First, you could say ``nothing.''  You can get by, and most people do, without ever explicitly paying attention to quantum physics at any time in your life.  


Second, you could say ``well, it is of practical importance to technology,'' because it is.  Whether you care or not, you use devices all the time that employ quantum physics.  Quantum theory is directly tied to applications with obvious economic value.  Many of those technologies, like nuclear weapons, also raise obvious ethical questions.  As an important current example, researchers are making significant advances in the development of computers that use quantum entanglement for novel types of computations and improvements in computing speed. Any serious consideration of the potential economic, social, or ethical impact of this new technology will require understanding  how quantum computation is distinct from classical computation, which means understanding some of the basics of quantum physics itself.  

Third, like scholar Karen Barad, you could say quantum physics changes ``everything,'' because it tells us that the universe does not respect the basic preconceptions about reality that we develop as inhabitants of the macroscopic realm.  Thus, perhaps our entire philosophical worldview, and even our vocabulary (which is normally quite bound in a thing-based ontology) should completely shift.  If, on a fundamental level, relationships are more definable than the things doing the relating, should that challenge how we view the concept of a relationship on any scale?  If, on a fundamental level, the properties of entities are indeterminate until interactions occur, should we give up any formal distinctions between subject and object in every context?  These are the kinds of philosophical leaps you might take if you commit to a certain interpretation of quantum physics and take quantum ontology as the final word.  

A fourth way that quantum physics might be valuable beyond the lab is as a source of metaphors and analogies. Quantum physics metaphors are rich. Opening up to thinking about non-thingness, indeterminate identity, blurry subject-object boundaries, and the dissolution of narrative may all be constructive things to do in our contemporary social and political moment, even if that context has little to do with the actual physics.  It might give us some new inspiration, and new points of view, for thinking differently. We want to underscore, though, that the association of these concepts with phenomena in physics does not confer any ``scientific'' authority to arguments that invoke quantum concepts metaphorically.  There is rarely a direct connection between the metaphorical context and the quantum physics context.  Moreover, we would argue that quantum concepts, expressed verbally, lack a certain kind of scientific authority in the first place, since we have no consensus on their interpretation.

There are also scholars who work with direct mathematical analogies.  This avoids some of the interpretation problems inherent in the verbal expression of quantum concepts. Yet, as with verbal metaphors, mathematical analogies based on quantum physics need not depend on (nor imply) any direct relationship to quantum physics itself. Quantum theory invokes a particular type of mathematical model (involving something called Hilbert spaces) to describe quanta. Similar mathematical models may be useful in other domains, such as cognition or finance\footnote{For an introduction to the field of quantum cognition, see the book   {\em{Quantum models of cognition and decision}}, by Jerome R. Busemeyer and Peter D. Bruza (Cambridge University Press, 2012).  For an example of analogies in the financial realm, see Schaden, Martin. "Quantum finance." {\em{Physica A: Statistical Mechanics and its Applications}} 316.1-4 (2002): 511-538.}.  But, consider that we can use a linear equation to describe the trajectory of an asteroid through space, or to describe the growth of savings in a child's allowance jar.  The fact that the same equation works in both cases does not require or imply any connection between asteroids and allowances.  Likewise, even though the name ``quantum'' is often used to label Hilbert space models, there need not be any connection between the non-physics and physics applications of such models.



In summary, there are many different ways that quantum physics might matter beyond the lab, perhaps as many as there are individual people who take the time to think about the question.  As your authors, we are interested in the ethics of quantum technology and curious about the potential of quantum metaphors and analogies.  We do not go as far as someone like Karen Barad, but we do take the stance that quantum physics presents productive and important challenges to person's worldview.  To us, the most important lessons are in the way that quantum physics requires {\em unlearning} ideas about how the world itself works, and about how scientific knowledge functions, that are grounded in macroscopic experience.  Because there is no single clear ontology implied by quantum physics, nature does not give us any solid or satisfying replacements for the naive ideas we are forced to give up.  

This is humbling, in a way that offers a counterbalance to some of the posturing we often observe from scientists.  Too often scientists and science communicators adopt the role of an authority full of answers, leading them, in the quantum physics case, to sweep problems of interpretation under the rug.  Individuals invoking quantum metaphors often seem eager to borrow this authoritative posture.  Yet, in our view the unsettled interpretation of quantum physics -- our persistent stuckness, our lack of authoritative answers -- is one of the most important things about it.  Even though quantum physics has unquestionably expanded our knowledge of the world, it also forces us to consider that some knowledge may be impossible.  Here we stand a full century after the development of quantum physics, and yet we are arguably no closer to resolving basic questions about {\em what is} in this universe we inhabit.



As a final note, we recently described this essay and the notion of ``obliterating thingness'' to School of the Art Institute of Chicago grad student Joshi Radin.  In response, she immediately quipped ``obliterating thingness sounds like an experience you can get on drugs. Why do you need quantum physics?''

It was a good point, so we just want to reiterate: non-thingness as a central feature of quantum physics is not about ``everything affects everything else'' or a breakdown of physical barriers in the world.  It is not something that, for us, translates to the head-space of feeling oneness with the universe or peace or comfort.  That's quantum ``woo'' talking, not quantum physics.  The way that quantum physics obliterates thingness is in the way that it undermines our ability to use language, and the thought structures associated with it (like narrative), to label and describe what we observe in nature when we test its behavior on small scales. It has more to do with a breakdown of our ability to represent reality in ways that feel like they make intuitive sense, leaving us with equations and recipes but no clear understanding of what they actually mean.  

Again, to us, the feeling that comes along with contemplating quantum physics is nothing like a sense of peace or wholeness or connectedness.  It is the feeling of deep humility, often tinged with frustration.  It does not matter how many years you spend as an expert in quantum physics, how much confidence you have in the project of science, or how hard you try to make quantum physics make sense.  You will often still find yourself wanting to scream ``what the fuck, universe?'' while staring at even the most basic experimental results.  Quantum reality deeply undermines the sense-making processes we are used to being able to perform as humans. A hundred years of effort has yet shown no way out of the fog.  It is possible that the fog is permanent. And that is deeply, deeply humbling, to us, even as we still experience wonder in the power of scientific inquiry.  That deep humility is something we do hope to share more broadly.  It is a form of cultural processing that we personally value and hope to add to the rich interdisciplinary conversations underway about what all of this means.

\section{Further Reading}

Our intent is for this essay to complement other available readings out there that introduce quantum physics in non-technical language.  Thus, we have deliberately avoided going through many of the pedagogical examples (the well-known ``double slit experiment'' for example) that often anchor those introductions.  We have also deliberately avoided some of the common vocabulary -- like wave-particle duality -- because we want to highlight that such vocabulary choices are linked to specific quantum interpretations.  Keep these points in mind to make the best use of this essay in conjunction with some of the readings below, or others that you find on your own.  Also look through the footnotes for some additional references.

\begin{enumerate}\item Raymer, Michael. "Quantum Physics: What Everyone Needs to Know", Oxford University Press (2017).\par
All the basic elements of quantum physics, including some potential applications, explained to non-scientists in a precise, yet simple and pedagogical text. This is a great first encounter with quantum phenomenology.

\item Albert, David Z. "Quantum mechanics and experience'', Harvard University Press (1992).
\par A slightly technical exposition of the many interpretations of quantum mechanics and their limitations. This book is not meant as a first encounter with quantum physics. It is a  good book for those who have already a basic understanding of quantum phenomena, and want to dig into their different philosophical interpretations.
\item Whitaker, A. "Einstein, Bohr and the Quantum Dilemma", Cambridge University Press (1996).
\par A very detailed account of the development of quantum theory, focusing on its history and its philosophy.
\item Barad, K. "Meeting the universe halfway: Quantum Physics and the Entanglement of Matter and Meaning" , Duke University Press Books (2007).
\par An intriguing non-technical book in which quantum phyics is connected to science studies, feminist, poststructuralist, and other critical social theories.
\end{enumerate}

\textbf{Acknowledgements:} This essay grew from a series of conversations that were made possible by the Scientist in Residence program at the School of the Art Institute of Chicago (SAIC), which brought Gabriela Barreto Lemos to the SAIC campus for the Fall semester of 2016.  We also acknowledge the support of the SAIC Undergraduate Dean's office and the SAIC Department of Liberal Arts as the primary sponsors of a Spring 2018 symposium entitled ``Quantum Unlearning,'' for which an earlier version of this essay was prepared to provide background reading.  Artist (and symposium co-organizer) Kyle Bellucci Johanson has shaped the direction of this essay significantly and provided invaluable feedback.  We have benefited from conversations with individuals spanning many disciplines, among which we wish to specifically thank Jacques Pienaar, Erik Nichols, Joseph Kramer, and Robb Drinkwater for helpful comments on prior drafts.  We thank the students in Kathryn's Spring 2018 ``Matter, Deconstructed'' course at SAIC for providing useful feedback on a couple of confusing points.  Finally, we thank the organizers and participants involved in the Spring 2018 Symposium ``Quantum Theory and the International'' at the Mershon Center for International Studies at Ohio State University, which one of us (Schaffer) attended, and which ultimately spurred the revision and publication of this work. 



\end{document}